\documentclass[sn-mathphys-num]{sn-jnl}

\usepackage{graphicx}%
\usepackage{multirow}%
\usepackage{amsmath,amssymb,amsfonts}%
\usepackage{amsthm}%
\usepackage{mathrsfs}%
\usepackage[title]{appendix}%
\usepackage{xcolor}%
\usepackage{textcomp}%
\usepackage{manyfoot}%
\usepackage{booktabs}%
\usepackage{algorithm}%
\usepackage{algorithmicx}%
\usepackage{algpseudocode}%
\usepackage{listings}%
\usepackage{makecell}
\usepackage{tikz}
\usetikzlibrary{positioning} 
\usetikzlibrary{fit} 
\usetikzlibrary{shapes.geometric,shapes.arrows,shapes.symbols,decorations.pathreplacing} 
\usepackage{arydshln} 
\usepackage{bm} 

\makeatletter
\usepackage{comment}
\let\wfs@comment@comment\comment
\let\comment\@undefined
\usepackage{changes}
\@namedef{Changes@AuthorColor}{red}
\definechangesauthor[color=red]{A}
\makeatother

\newcommand{\rev}[1]{#1} 
\newcommand{\revII}[1]{#1}
\newcommand{\eat}[1]{}

\newcommand{\osnomed}{SNOMED-CT}
\newcommand{\oloinc}{LOINC}
\newcommand{\oomim}{OMIM}
\newcommand{\ohgnc}{HGNC}

\begin{document}

\title
{I-ETL: an interoperability-aware health (meta)data pipeline to enable federated analyses}


\author[1]{\fnm{Nelly} \sur{Barret}}\email{nelly.barret@polimi.it}

\author[1]{\fnm{Anna} \sur{Bernasconi}}\email{anna.bernasconi@polimi.it}

\author[1]{\fnm{Boris} \sur{Bikbov}}\email{boris.bikbov@polimi.it}

\author*[1]{\fnm{Pietro} \sur{Pinoli}}\email{pietro.pinoli@polimi.it}

\affil[1]{\orgname{Department of Electronics, Information and Bioeingineering -- Politecnico di Milano}, \orgaddress{\country{Italy}}}


\abstract{


\textbf{Background:}  Clinicians are interested in better understanding complex diseases, such as cancer or rare diseases, so they need to produce and exchange data to mutualize sources and join forces. \rev{To do so and ensure privacy, a natural way consists in using a decentralized architecture and Federated Learning algorithms. This ensures that data stays in the organization in which it has been collected, but requires data to be collected in similar settings and similar models. In practice, this is often not the case because healthcare institutions work individually with different representations and raw data; they do not have means to normalize their data, and even less to do so across centers. For instance, clinicians have at hand phenotypic, clinical, imaging and genomic data (each individually collected) and want to better understand some diseases by analyzing them together. This example highlights the needs and challenges for} a cooperative use of this wealth of information.

\textbf{Methods:} We designed and implemented a framework, named I-ETL, for integrating highly heterogeneous healthcare datasets of hospital\rev{s} in interoperable database\rev{s}. Our proposal is twofold: ($i$)~we devise two general and extensible conceptual models for modeling both data and metadata and ($ii$)~we propose an Extract-Transform-Load (ETL) pipeline ensuring and assessing interoperability from the start. 

\textbf{Results:} \rev{By} conducting experiments on open-source datasets, we show that I-ETL succeeds in representing various health datasets in a unified way thanks to our two general conceptual models. Next, we demonstrate the importance of blending interoperability as a first-class citizen in integration pipelines, ensuring possible collaboration between different centers. 

\textbf{Conclusions:} As a framework, I-ETL contributes to integrat\rev{e} and improv\rev{e} interoperability between healthcare institutions. \rev{When used in a decentralized federated platform, it eases the federated analysis of the different hospital databases and} helps clinicians to \rev{obtain insights and knowledge on} medical conditions of interest.  
}


\maketitle



\section{Background}
\label{sec:background}

The overall recent digitization of the healthcare sector has led to new opportunities for researchers and clinicians to access, integrate, analyze, share, and reuse medical data~\cite{big-data-health}. \rev{An example of this is the emergence of self-tracking tools, 
which are source of plenty of healthcare data for various profiles of citizens~\cite{bietz2016opportunities}.} Plenty of \rev{other} initiatives have been conducted and deployed at different levels, from city- \rev{and} national-\rev{, to} European\rev{-} \rev{and world}-wide platforms, and on diverse healthcare major open issues (access to healthcare, cancer, genetic rare diseases, etc). For instance, \rev{the World Health Organization (WHO) collects world-wide data in their data hub~\cite{who-data} on various healthcare domains, including COVID-19, mortality and access to healthcare, and computes yearly reports~\cite{who-stats} after integrating and analyzing it. Another example is} the European Union\rev{, which} heavily invest\rev{s} in rare diseases investigation~\cite{eu-rare-diseases} since 2007, notably to develop new tools to decipher them, as they remain largely poorly understood. As of today, around 6,000 rare diseases are known and 80\% of them are of genetic origin, most of which have no effective treatment or allow for easy diagnosis. 
Conducting data-driven research in the context of rare diseases is evidently challenging, as assembling a sufficiently large dataset would require aggregating medical and genetic data from patients across numerous geographically distributed clinical institutions; however, current regulations, such as the General Data Protection Regulation (GDPR), impede the aggregation of sensitive data in a centralized repository. In response to this, the scientific community is developing tools and solutions to analyze medical and genetic data in a federated and secure manner. These tools do not require the exchange of real data but solely aggregate extracted information (e.g., statistics or partial models) computed from local datasets. Performing federated analyses allows the study of larger sets of patients, potentially with heterogeneous types of data, thus obtaining more accurate results and insights.

However, from a data perspective, federated analyses require a significant effort in standardization and harmonization to ensure interoperability among distributed datasets. It is clear that manual curation and cleaning are not feasible or scalable. Therefore, collaborating hospitals crucially need tools to automatically integrate heterogeneous and sensitive datasets with interoperability as a first-class citizen.





\subsection{\rev{Conceptual models for healthcare data}}
Several large projects have proposed instruments for modeling and enforcing interoperability of distributed heterogeneous healthcare datasets, intending to facilitate federated analyses. Notably, the {EHDEN project}~\cite{ehden,ehdenII} safely integrates hundreds of relational (tabular) datasets across Europe to provide clinicians a single endpoint where they can select datasets of interest for observational studies, define protocol\rev{s} and run analyses. For unifying the different concepts across tables, they used the \textit{OMOP}~\cite{omop} (Observational Medical Outcomes Partnership) conceptual model as a Common Data Model (CDM). \rev{Similarly,~\cite{revsvcivc2023smartchange} converts European healthcare data to the FHIR~\cite{fhir} data model in order to define Deep Learning models for early-diagnosis of children and teenagers regarding non-communicable chronic diseases.} 
There exist many widely-adopted CDMs with the same intent of \textit{OMOP} \rev{and \textit{FHIR}}; the most significant are \rev{i2b2~\cite{i2b2} for precision medicine}, \textit{GCM}~\cite{anna-genomic-model,bernasconi2022comprehensive} for genomic data (developed within the GeCo ERC AdG project~\cite{ceri2017overview}) \rev{and} the \textit{Human Cell Atlas}~\cite{human-atlas} for human cells, but also \rev{general} ones such as \textit{openEHR}~\cite{openehr}, and \textit{SMART}~\cite{smart}.  

To fit the input data into a CDM of choice, it is common to rely on an ETL pipeline (Extract-Transform-Load). This is a three-step process where data is extracted from input sources, transformed/cleaned, and loaded into a target data container, usually a database, whose schema is \rev{the} CDM. D-ETL~\cite{detl} is a dynamic ETL pipeline partially automating the process by providing data harmonization techniques and simplifying the transformation process. In turn, experts specify ``ETL structured rules'' for mapping the actual input data to the CDM. Then, these rules are transformed into SQL statements and data is loaded into the target database. \rev{For the more general FHIR standard,~\cite{kouremenou2023data} proposes a 6-step workflow to help healthcare scientists to model their data with FHIR entities -- this is a more conceptual approach of the ETL process.} 


\subsection{Metadata for better interoperability} 
To allow interoperability among datasets situated at different institutions\rev{,} it is crucial to accurately describe each piece of information using unambiguous metadata. 
\textbf{Metadata} is supplementary information that allows the assignment of meaning to both the type of data and its associated value. 
While metadata can be of multiple forms, in clinical and biomedical settings it is a best practice to (re)use existing ontologies that describe data concepts. Many specialized ontologies exist, such as \osnomed~\cite{snomed} for general-purpose healthcare terms, \oloinc~\cite{loinc,loincII} for clinical measurements, or OrphaNet~\cite{orphanet} for disease classification. By using ontologies, data points can be mapped to unique ontolog\rev{y resources}, thus enabling interoperability among datasets \rev{of interest}.

Moreover, metadata are essential to interoperability, which is part of the ``I" prescription of the FAIR principles~\cite{fair-wilkinson}, a set of guidelines to make data and its accompanying metadata Findable, Accessible, Interoperable, and Reusable. 
In general, FAIR principles encourage IT experts to identify their resources with unique and reusable identifiers (e.g., URIs), use widely used standards and protocols, and provide rich metadata by reusing existing ontologies. 
They are, by definition, general enough to allow their adoption by very heterogeneous projects. Therefore, many healthcare integration systems have been designed with FAIR principles in mind; this is the case of UMG-MeDIC~\cite{fairness-intensive-care} and {Scaleus-FD}~\cite{scaleus}. 

Although the principles of FAIRness are applied during dataset processing, it is also important to assess whether the dataset (or other digital object, such as a database) itself meets the FAIR criteria. According to a recent survey~\cite{fair-metrics-survey}, existing assessment tools are often tied to a given context, may involve manual assessment, and often focus solely on data FAIRness, not considering metadata and semantics FAIRness.
A possible solution to overcome these limitations is to integrate FAIRness assessment from the start and provide explainable metrics to users, to let them improve the quality (and FAIRness) of their data and metadata. 

\subsection{Limitations and contributions} 

\rev{As illustrated before, many approaches have been proposed to model healthcare datasets that are very heterogeneous (in their model and content) and} to enforce interoperability\rev{. Nevertheless, they} exhibit some limitations. 

First, many of them are tied to a single data model (e.g., EHDEN~\cite{ehden, ehdenII} only integrates tabular data) or leverage a CDM tied to a\rev{n healthcare domain} (e.g., OMOP~\cite{omop} models observational data only \revII{and is hardly extensible to other domains like genomic or imaging data}). \revII{Second, they all} exhibit entities with specific attributes \revII{(see, for instance, the FHIR \textit{Observation} entity with its 24 attributes, including \texttt{reason}, \texttt{status}, \texttt{subject}, and \texttt{value}). In turn, these models} lead to hand-made integration workflows (to map data concepts with those in the model \revII{and to adapt to specific attributes}) or very abstract workflows. \revII{As an example,} Dynamic-ETL~\cite{detl} \revII{(a data integration pipeline)} requires experts to write rules expressing how to match source models to OMOP. \revII{On the contrary,~\cite{kouremenou2023data} proposes a FHIR-based workflow to transform existing medical data to FHIR. However, it} do\revII{es} not include practical steps to realize the data integration part. Th\revII{e above reasons and examples emphasize why existing models are not convenient for the transformation of existing data to those models (while we note that they are suited for collecting new data). The main limitations include the attribute-based models leading to specific pipelines that are hardly reusable, and the technical barriers (e.g., map input and target models, and write corresponding data rules) that experts may not be able to surpass.} 
\rev{Finally,} FAIR metrics \rev{must be computed} along ETL pipelines to ensure high findability, accessibility, interoperability, and reusability\revII{, while most works focus on assessing the interoperability on the obtained data~\cite{fair-metrics-survey}}. 

In this work, we propose \rev{\textbf{I-ETL}}, a novel framework \rev{to} enforc\rev{e} interoperability among heterogeneous distributed healthcare datasets. \rev{It} ensures privacy, requires moderate input from clinical experts, and computes a holistic interoperability assessment. Our main contributions are: 
\begin{enumerate}
    \item Two tightly linked, extensible, \textbf{conceptual models for both metadata and data}, based on experts' knowledge, for achieving data interoperability within and across hospitals' data stores.
    \item An \textbf{ETL pipeline} in which interoperability is a first-class citizen, producing a target database to be used in federated and distributed analytics contexts.
    \item Guarantee of \textbf{interoperability} using a set of metrics that are progressively assessed during I-ETL.
\end{enumerate}


\section{Methods}
\label{sec:methods}

\subsection{I-ETL approach and concepts} 
\label{ssec:pipeline}

\begin{figure}
    \centering
    \resizebox{\columnwidth}{!}{
    \begin{tikzpicture}[node distance=1cm and 0cm, inner sep=2pt]
        \node (t2) [shape=signal, signal from=west, draw, align=center] {select\\a set of\\datasets};
        \node (t3) [shape=signal, signal from=west, draw, align=center, right=of t2] {define\\relevant\\features};
        \node (t4) [shape=signal, signal from=west, draw, align=center, right=of t3] {fill\\the\\metadata};
        \node (t5) [shape=signal, signal from=west, draw, align=center, right=of t4] {apply\\the\\ETL};
        \node (t6) [shape=signal, signal from=west, draw, align=center, right=of t5] {yield\\interoperability\\metrics};
        
        \node (Dn) [above=of t2] {$D_1, ..., D_n$};
        \node (F) [above=of t3] {$F$};
        \node (metadata) [above=of t4] {$M$};
        \node (db) [cylinder, draw, shape border rotate=90, aspect=0.05, above=of t5, align=center,yshift=-5mm] {Database with\\CDM as schema};
        \node (metrics) [above=of t6, draw=none, align=center] {report};

        \node (leftarrow) [below=of t3, xshift=-12mm] {};
        \node (rightarrow) [below=of t5,xshift=10mm] {};

        
        \node (m2) [below=of t3, align=center, yshift=8mm] {data\\metrics};
        \node (m4) [below=of t4, align=center, yshift=8mm] {metadata\\metrics};
        \node (m5) [below=of t5, align=center, yshift=8mm] {ETL\\metrics};

        \draw [->,dashed] (t2) -> (Dn);
        \draw [->,dashed] (t3) -> (F);
        \draw [->,dashed] (t4) -> (metadata);
        \draw [->,dashed] (t5) -> (db);
        \draw [->,dashed] (t6) -> (metrics);
        
        \draw [->, line width=1pt] (leftarrow) -> (rightarrow) node [midway,below,sloped,draw=none,inner sep=1.5mm] {Across-pipeline interoperability metrics};
    \end{tikzpicture}
}
    \caption{I-ETL, the framework building interoperable databases for federated analyses from heterogeneous healthcare data. Large arrows represent steps in the framework; dashed edges connect a step to its corresponding (intermediate) result. Interoperability metrics are shown below the task during which they are computed. 
    \label{fig:pipeline}}
\end{figure}
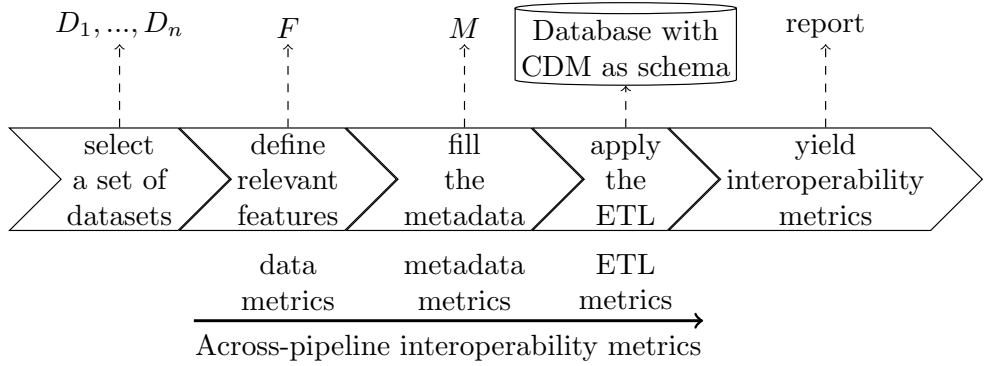

The proposed I-ETL approach is a 5-step data science pipeline (see Figure~\ref{fig:pipeline}). Starting from the left side, medical experts decide collaboratively on a specific topic to study, e.g., pediatric intelligence disability or kidney cancer, and \textit{select a set of datasets} relevant to their chosen topic. 
A \textbf{dataset} is any sort of file containing data about patients;
it may be, for instance, 
a CSV file containing the clinical measures obtained from blood samples, 
a DICOM file obtained from an eye MRI scan, 
a VCF file comprising patients' genomic variants, etc. 

Next, after inspecting the selected datasets, practitioners \textit{define a set of relevant features} $F$ in those datasets. 
In this work, a \textbf{feature} is a specific attribute of data, e.g., the birth date of a patient, the size of the dark regions in MRI eye scans, etc. 

Keeping $F$ in mind, practitioners \textit{fill the metadata} $M$ of the chosen datasets. We define \textbf{metadata} as any information providing the {context} to understand and interpret a feature. Metadata typically includes the feature name, its data type, possibly an ontology resource that could be associated with it, etc. (details on our metadata model are given in Section~\ref{ssec:metadata-model}). 

The fourth step in the global pipeline is the ETL 
process (later described in Section~\ref{ssec:etl}). In a nutshell, this aims at transforming the input datasets into a target database whose schema is our general conceptual model for healthcare datasets (presented in Section~\ref{ssec:data-model}), enabling interoperability between the ingested datasets, but also with the other databases used in the federated analysis task.
During the pipeline, interoperability metrics are collected (see Section~\ref{ssec:interop-metrics}) and are then reported to practitioners.

\subsection{Metadata model}
\label{ssec:metadata-model}

After picking relevant datasets for a use case of interest, practitioners have to select or extract a set of relevant features in their datasets. However, simply selecting (or extracting) a set of relevant features is not sufficient because individual datasets, or hospitals, may encode similar features with different names, thus prohibiting interoperability. To overcome this semantic issue, we ask experts to specify which ontology resource may be used to represent each feature. \rev{This mapping of features to existing ontology resources, done during the metadata creation, improves interoperability and alignment between hospitals. Unmapped features can still be referred to by their original names, even though this considerably hinders interoperability.} 

The following list depicts our one-entity \textbf{metadata model} for describing datasets with interoperability as a first-class citizen: 
\begin{itemize}
    \item \textbf{Name}: the name of the feature;
    \item \textbf{Ontology}: the name of the ontology chosen to represent the feature concept where ontologies can be selected in well-known portals like BioPortal~\cite{bioportal-ontos};
    \item \textbf{Code}: the code of the resource in the selected ontology that represents the feature;
    \item \textbf{Kind}: the type of feature, i.e., phenotypic, clinical, genomic, image, etc;
    \item \textbf{DataType}: the {expected} \rev{value} type (among \texttt{string}, \texttt{integer}, \texttt{numeric}, \texttt{boolean}, \texttt{category}, \texttt{date}, or \texttt{datetime});
    \item \textbf{Unit}: the unit to interpret the values when the feature data type is integer or numeric;
    \item \textbf{Categories}: when the feature is categorical, each value is paired to an ontology \rev{resource (a pair of the ontology name and a code)}.
    \item \textbf{Visibility}: whether the values for that feature can be shown publicly (\texttt{public}), after anonymization (\texttt{anonymized}), or cannot be shown at all (\texttt{private}), depending on their sensitivity.
\end{itemize}


~\newline
\noindent\textbf{Example instances.} 
Assume we have two phenotypic features selected from datasets for the kidney disease scenario, namely \rev{age} and \rev{sex}:
\begin{itemize}
\item The first feature is instantiated with the following values: \textit{name} is ``\rev{calc\_age}'', \textit{ontology} and \textit{code} are \osnomed\ and ``397669002'',  \textit{dataType} is \texttt{integer}, \textit{unit} is ``years'' \rev{while} \textit{categories} is null (the feature is not categorical), and the \textit{visibility} is \texttt{anonymized} to prevent the access to the original patient \rev{age} (the \rev{age} could be deduced if the patient is younger or older than the average). 

\item The second feature, about the \rev{sex}, is instantiated with \textit{name} ``\rev{sex}'',   \textit{ontology} \rev{is} \osnomed\ \rev{with the \textit{code} ``734000001''}. The \textit{dataType} is \texttt{category}, there will be no \textit{unit} and the \textit{visibility} would be \texttt{public}. For \textit{categories}, we have \rev{two} pairs: \rev{\texttt{(Female, (\osnomed, 248152002))} \rev{and} \texttt{(Male, (\osnomed, 248153007))}.}
\end{itemize}

\subsection{Common data model}
\label{ssec:data-model}

After describing features with metadata, we run the ETL pipeline to transform the input datasets in a homogenized, interoperable target database. This relies on a \textbf{common data model} (CDM in short), which is a conceptual model for representing homogeneously and making all the selected data interoperable. Each database included in a federated scenario, 
implements the same CDM so that all databases can cooperate, be accessed, and queried in a unified way, regardless of the heterogeneous datasets they carried in origin.

\begin{figure}
    \centering
    \tikzstyle{myNode} = [inner sep=0mm]
\tikzstyle{relNode} = [draw, rectangle, rounded corners]
\tikzstyle{myTriangle} = [regular polygon, regular polygon sides=3]
\tikzstyle{collEdgeLabelHL} = [near start, draw=none, above, xshift=1mm]
\tikzstyle{collEdgeLabelHR} = [near start, draw=none, above, xshift=-1mm]
\tikzstyle{collEdgeLabelV} = [near start, draw=none,xshift=3mm]

\resizebox{\columnwidth}{!}{
\begin{tikzpicture}[node distance=0.6cm and 0.5cm]
    \node (Record) [myNode] {\begin{tabular}{|l|l|} \hline \multicolumn{2}{|c|}{Record} \\ \hline \underline{identifier} & ID \\ \hline value & any \\ \hline dataset & str \\ \hline \end{tabular}};
    
    \node (recBy) [relNode, left=of Record,xshift=-1mm] {register};
    
    \node (Hospital) [myNode, left=of recBy,xshift=-1mm] {\begin{tabular}{|l|l|} \hline \multicolumn{2}{|c|}{Hospital} \\\hline \underline{identifier} & ID \\ \hline name & str \\ \hline \end{tabular}};
    
    \node (subject) [relNode, right=of Record,xshift=1mm] {associate};
    
    \node (Patient) [myNode, right=of subject,xshift=1mm] {\begin{tabular}{|l|l|} \hline \multicolumn{2}{|c|}{Patient} \\\hline \underline{identifier} & ID \\ \hline  \end{tabular}};

    \node (trRec) [draw, myTriangle, above=of Record,yshift=-3mm] {};

    \node (DiagnosisRec) [myNode, above=of trRec] {\begin{tabular}{|c|} \hline Diagnosis \\ Record \\\hline \end{tabular}};
    \node (MedicineRec) [myNode, left=of DiagnosisRec,xshift=3mm] {\begin{tabular}{|c|} \hline Medicine \\ Record \\\hline \end{tabular}};
    \node (ClinicalRec) [myNode, left=of MedicineRec,xshift=3mm,yshift=-2mm] {\begin{tabular}{|l|l|} \hline\multicolumn{2}{|c|}{Clinical} \\ \multicolumn{2}{|c|}{Record} \\\hline baseId* & ID \\ \hline \end{tabular}};
    \node (PhenotypicRec) [myNode, left=of ClinicalRec,xshift=3mm,yshift=2mm] {\begin{tabular}{|c|} \hline Phenotypic \\ Record \\\hline \end{tabular}};
    \node (GenomicRec) [myNode, right=of DiagnosisRec,xshift=-3mm,yshift=-2mm] {\begin{tabular}{|l|l|} \hline \multicolumn{2}{|c|}{Genomic} \\ \multicolumn{2}{|c|}{Record} \\\hline vcf* & uri \\ \hline \end{tabular}};
    \node (ImagingRec) [myNode, right=of GenomicRec, xshift=-3mm] {\begin{tabular}{|l|l|} \hline \multicolumn{2}{|c|}{Imaging} \\ \multicolumn{2}{|c|}{Record} \\\hline scan* & uri \\ \hline \end{tabular}};
    \node (XRec) [myNode, right=of ImagingRec, xshift=-2mm, yshift=2mm] {\begin{tabular}{:c:} \hdashline ... \\ Record \\\hdashline\end{tabular}};
    \node (instPhen) [relNode, above=of PhenotypicRec] {instantiate};
    \node (instClin) [relNode, above=of ClinicalRec] {instantiate};
    \node (instMed) [relNode, ,above=of MedicineRec] {instantiate};
    \node (instDiag) [relNode, above=of DiagnosisRec] {instantiate};
    \node (instGen) [relNode, above=of GenomicRec] {instantiate};
    \node (instImg) [relNode, above=of ImagingRec] {instantiate};
    \node [dashed] (instX) [relNode, above=of XRec] {instantiate};
    \node (PhenotypicFeat) [myNode, above=of instPhen] {\begin{tabular}{|c|} \hline Phenotypic \\ Feature \\\hline \end{tabular}};
    \node (ClinicalFeat) [myNode, above=of instClin] {\begin{tabular}{|c|} \hline Clinical \\ Feature \\\hline \end{tabular}};
    \node (MedicineFeat) [myNode, above=of instMed] {\begin{tabular}{|c|} \hline Medicine \\ Feature \\\hline \end{tabular}};
    \node (DiagnosisFeat) [myNode, above=of instDiag] {\begin{tabular}{|c|} \hline Diagnosis \\ Feature \\\hline \end{tabular}};
    \node (GenomicFeat) [myNode, above=of instGen] {\begin{tabular}{|c|} \hline Genomic \\ Feature \\\hline \end{tabular}};
    \node (ImagingFeat) [myNode, above=of instImg] {\begin{tabular}{|c|} \hline Imaging \\ Feature \\\hline \end{tabular}};
    \node (XFeat) [myNode, above=of instX] {\begin{tabular}{:c:} \hdashline ... \\ Feature \\\hdashline \end{tabular}};

    \node (trFeat) [draw, myTriangle, above=of MedicineFeat] {};

    \node (Feature) [myNode, above=of trFeat,yshift=-5mm] {\begin{tabular}{|l|l|} \hline \multicolumn{2}{|c|}{Feature} \\\hline \underline{identifier} & ID \\ \hline name & str \\ \hline dataType* & str \\ \hline unit* & str \\ \hline categories* & map \\ \hline visibility* & str \\ \hline \end{tabular}};
    
    \node (repBy) [relNode, right=of Feature,xshift=2mm] {represent};
    
    \node (OntoRes) [myNode, right=of repBy,xshift=2mm] {\begin{tabular}{|l|l|} \hline \multicolumn{2}{|c|}{Ontology} \\ \multicolumn{2}{|c|}{Resource} \\\hline \underline{system} & str \\ \hline \underline{code} & str \\ \hline label* & str \\ \hline \end{tabular}};

    \draw (Record) -> (recBy) node[collEdgeLabelHR] {0..n};
    \draw (Hospital) -> (recBy) node[collEdgeLabelHL] {1..1};
    \draw (Record) -> (subject) node[collEdgeLabelHL] {0..n};
    \draw (Patient) -> (subject) node[collEdgeLabelHR] {1..1};
    \draw (trRec) -> (Record);
    \draw (PhenotypicRec) |- (trRec) ;
    \draw (PhenotypicRec) -> (instPhen) node[collEdgeLabelV] {0..n};
    \draw (ClinicalRec) |- (trRec);
    \draw (ClinicalRec) -> (instClin) node[collEdgeLabelV] {0..n};
    \draw (MedicineRec) -> (trRec);
    \draw (MedicineRec) -> (instMed) node[collEdgeLabelV] {0..n};
    \draw (DiagnosisRec) -> (trRec);
    \draw (DiagnosisRec) -> (instDiag) node[collEdgeLabelV] {0..n};
    \draw (GenomicRec) -> (trRec);
    \draw (GenomicRec) -> (instGen) node[collEdgeLabelV] {0..n};
    \draw [yshift=2mm] (ImagingRec) |- (trRec);
    \draw (ImagingRec) -> (instImg) node[collEdgeLabelV] {0..n};
    \draw [dashed] (XRec) |- (trRec);
    \draw [dashed] (XRec) -> (instX) node[collEdgeLabelV] {0..n};
    \draw (PhenotypicFeat) |- (trFeat);
    \draw (PhenotypicFeat) -> (instPhen) node[collEdgeLabelV] {1..1};
    \draw (ClinicalFeat) -> (trFeat);
    \draw (ClinicalFeat) -> (instClin) node[collEdgeLabelV] {1..1};
    \draw (MedicineFeat) -> (trFeat);
    \draw (MedicineFeat) -> (instMed) node[collEdgeLabelV] {1..1};
    \draw (DiagnosisFeat) -> (trFeat);
    \draw (DiagnosisFeat) -> (instDiag) node[collEdgeLabelV] {1..1};
    \draw (GenomicFeat) |- (trFeat);
    \draw (GenomicFeat) -> (instGen) node[collEdgeLabelV] {1..1};
    \draw (ImagingFeat) |- (trFeat);
    \draw (ImagingFeat) -> (instImg) node[collEdgeLabelV] {1..1};
    \draw [dashed] (XFeat) |- (trFeat);
    \draw [dashed] (XFeat) -> (instX) node[collEdgeLabelV] {1..1};
    \draw (trFeat) -> (Feature);
    \draw (Feature) -> (repBy) node[collEdgeLabelHL] {0..n};
    \draw (OntoRes) -> (repBy) node[collEdgeLabelHR] {0..1};
\end{tikzpicture}
}
    \caption{The interoperable conceptual model instantiated at each medical center database. Rectangles are entities, rounded boxes are relationships and triangles are specializations. Primary keys are \underline{underlined}, and optional attributes are marked with a * (star).
    Our cardinalities adopt the notation in~\cite{chen1976entity}, e.g., a Record instantiates exactly one Feature, is associated with exactly one Patient, and is registered by exactly one Hospital. Features can be instantiated in 0 to n Record entities.}
    \label{fig:conceptual-model}
\end{figure}
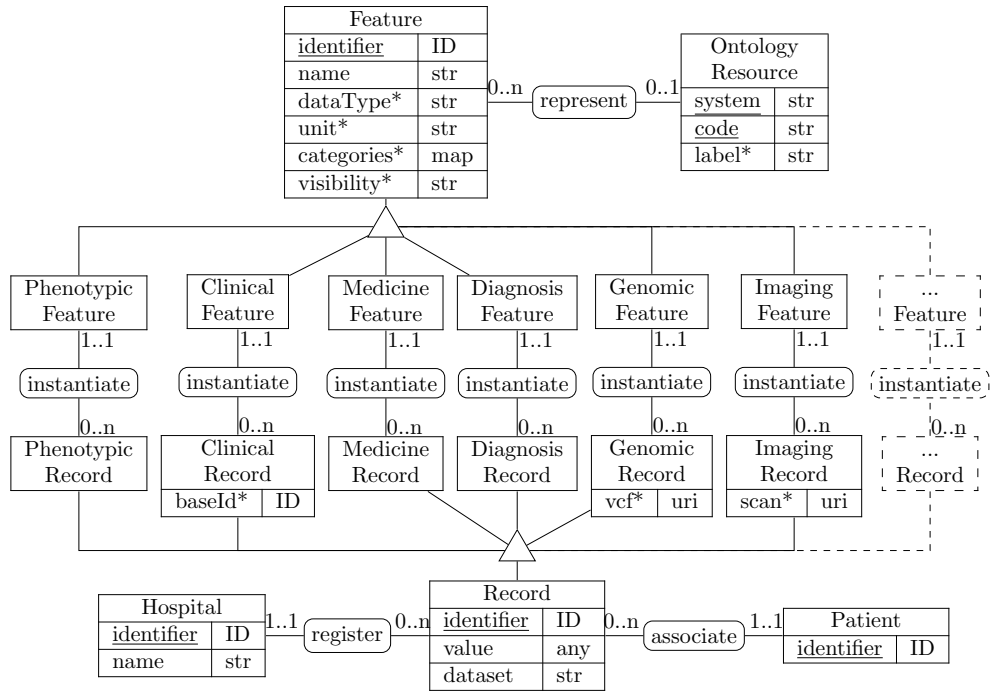

Our CDM is illustrated in Figure~\ref{fig:conceptual-model}; it promotes two important characteristics. First, it isolates medical knowledge and concepts specified in the metadata from the actual data instances by distinguishing two concepts: \textbf{features and records}. \rev{These two concepts are abstractions of how any dataset can be represented and make our model a common data model (as opposed to a project-based data model).} 
As previously defined, a \textsc{Feature} is an attribute of the data, e.g., the birth date of a patient. 
Instead, a \textsc{Record} is the actual value that a patient has for a certain feature, as produced by a\rev{n} hospital.
For instance, according to hospital H1,
for patient P1 and a feature whose \textit{name} is ``birth\_date'' the value is ``01/01/2000''. 
Second, it describes six different kinds of data (phenotypic, clinical, medicine, diagnosis, genomic, and imaging), while being easily extensible to new types of data, e.g., administrative data or patient history. \rev{This also contributes to make our model general enough to be considered as a common data model, possibly reused by many other diverse projects whose goals are to model very heterogeneous healthcare datasets in various settings.}



In our conceptual model (Figure~\ref{fig:conceptual-model}), we follow the typical notation of E\rev{ntity-}R\rev{elationship} diagrams~\cite{chen1976entity} established in~\cite{batini1992conceptual}.
The central entity is the \textsc{Record}; each record has a unique \textit{identifier}, a \textit{value}, and the name of the \textit{dataset} to which it belongs. The \textit{value} is of type \texttt{any} as records may capture not only atomic values, such as integer or boolean, but also complex ones, e.g., categories. 
The \textit{value} can be \rev{further} anonymized depending on its sensitivity (see Section~\ref{ssec:etl}). \rev{The attribute \textit{dataset} allows to keep track of the provenance, i.e., from where individual values come from.} 

Since clinicians need various kinds of data when studying a research matter, the \rev{\textsc{Record}} entity is specialized in 6 entities, namely phenotypic, clinical, medicine, diagnosis, genomic, or imaging, each of which inherits the \rev{\textsc{Record}} attributes. 
Note that this set of data kinds can be easily extended \rev{or modified} for different scenarios. 

When a patient has an appointment, practitioners first collect phenotypic data, i.e., information about the patient, the environment and habits, each value leading to a \textsc{PhenotypicRecord}. 
Next, patients usually go through a series of tests in laboratories, including blood tests, breathing tests, skin tests, etc. For each test, captured values become \textsc{ClinicalRecord} \rev{instances}. If the clinical record is associated with a sample, the sample identifier is stored in the optional attribute \textit{baseId}. 
Genome sequences of patients are aligned on a default genome to obtain genomic variants. Those variants and their associated information, such as the chromosome on which it appears or the confidence\rev{,} are represented \rev{with the} \textsc{GenomicRecord} \rev{entity}. The genomic record also saves the VCF file path from which the value has been extracted with the attribute \textit{vcf}. 
Moreover, some patients are also asked to go for MRI scans to better visualize areas of interest. 
Each MRI scan is an image, from which a set of features are extracted and each value is stored as an \textsc{ImagingRecord}. As for genomic records, imaging records save the image file path from which the value has been extracted using the attribute \textit{scan}. 
For patients taking medicines, each information in a prescription is seen as a \textsc{MedicineRecord}. 
Finally, patients may obtain a diagnosis, leading to a \textsc{DiagnosisRecord}. 

Each \rev{\textsc{Record}} is registered by exactly one \textsc{Hospital}, i.e., a clinical institution contributing to the federated analysis task. It has an \textit{identifier} and a \textit{name}. A \rev{\textsc{Record}} has for subject exactly one \textsc{Patient}, i.e., an individual included in the studied cohort. Patients only have an \textit{identifier} because they are totally anonymized for privacy, thus, no further personal information can be included. 

Each \rev{\textsc{Record}} entity instantiates exactly one \textsc{Feature} of the same kind, e.g., a phenotypic record instantiates a phenotypic feature. Those features are specializations of the \textsc{Feature} entity, carrying all the attributes. Each feature has an \textit{identifier}; other attributes come from the metadata model. In turn, each feature has a \textit{name}, a \textit{dataType}, a \textit{unit}, a set of \textit{categories}, and a \textit{visibility}. The last four attributes are optional because: the \textit{dataType} and the \textit{visibility} may not be specified by medical experts (\textit{visibility} defaults to \texttt{private} for privacy reasons), the \textit{unit} exists only for numerical features, and the list of \textit{categories} only exists for categorical features. 

A \rev{\textsc{Feature}} is represented by zero or one \textsc{OntologyResource}. 
Sometimes, features cannot be formalized in any ontology (hence the cardinality \texttt{0..1}). 
It contains a \textit{system}, i.e., the endpoint URL to access the ontology, a \textit{code}, i.e., a unique identifier for the represented concept in that ontology, and a \textit{label}, a human-friendly name for that concept. 
All of them are strings, only \textit{system} and \textit{code} are mandatory (in order to identify the represented concept). 

A \textsc{PhenotypicFeature} captures any variable about the factors that may affect the patient, such as the environment, daily routine, habits, etc. 
A \textsc{ClinicalFeature} may be about any chemical or clinical measurement. 
A \textsc{GenomicFeature} typically concerns the chromosomes carrying variants, the exact position of the variant, the type of variation (addition, deletion, mutation), etc. 
\textsc{ImagingFeature} \rev{instances} can represent diverse features depending on the patient's disease. For instance, if the patient is affected by a brain tumor, the features may include the coordinates of the tumor in the MRI scan, the darkness of the tumor in the scan, etc. Additional features about the scanner and software can be included too, e.g., the software version, the scanner name, etc. 
A \textsc{MedicineFeature} captures variables such as the name of the medication, the start and length of the prescription, whether the patient took the medicine, etc. 
Finally, a \textsc{DiagnosisFeature} captures a characteristic of the diagnosis attributed to a patient. Diagnosis characteristics include the diagnosis name, but also the affected gene, whether the patient is a carrier or affected, etc. 


\subsection{ETL pipeline and target database}
\label{ssec:etl}

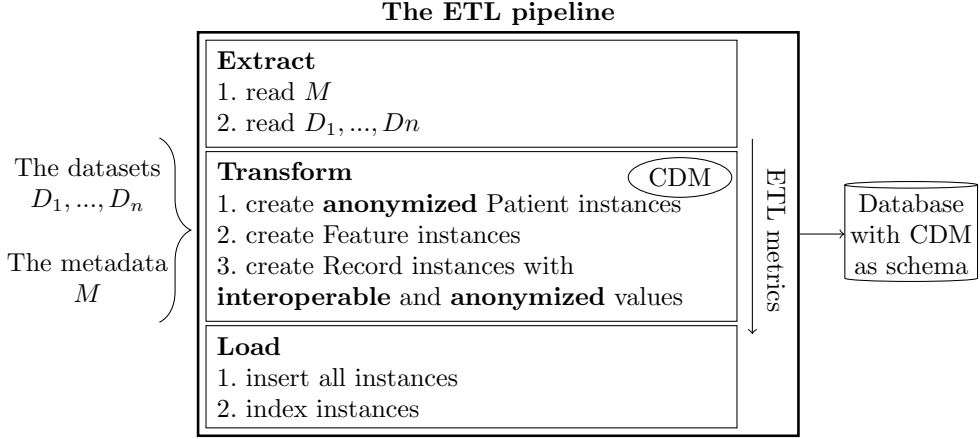
\begin{figure}
    \centering
    \resizebox{\columnwidth}{!}{
    \begin{tikzpicture}[node distance=0cm and 0.6cm, inner sep=2pt]
        \node (Extract) [rectangle, align=left,text width=68.5mm] {\textbf{Extract}\\1. read $M$\\2. read $D_1, ..., Dn$};
        \node (Transform) [rectangle, align=left, below=of Extract, yshift=-2mm,text width=68.5mm] {\textbf{Transform}\\1. create \textbf{anonymized} Patient instances\\2. create Feature instances\\3. create Record instances with \\\textbf{interoperable} and \textbf{anonymized} values};
        \node (Load) [rectangle, align=left, below=of Transform, yshift=-2mm,text width=68.5mm] {\textbf{Load}\\1. insert all instances\\2. index instances};
        
        \node (T1) [draw, fit=(Extract)] {};
        \node (T2) [draw, fit=(Transform)] {};
        \node (T3) [draw, fit=(Load) ] {};
        \node (m1) [right=of Transform, yshift=14mm, xshift=-4mm] {};
        \node (m2) [right=of Transform, yshift=-14mm, xshift=-4mm] {};
        \node (metrics) [right=of Transform, rotate=-90, xshift=-10mm] {ETL metrics};
        
        \node (I-ETL) [draw, fit=(T1) (T2) (T3) (metrics), label={\textbf{The ETL pipeline}}, inner sep=1mm, line width=1pt] {};
        
        \node (Dn) [align=center, left=of Transform,yshift=7mm] {The datasets\\$D_1, ..., D_n$};
        \node (metadata) [align=center, below=of Dn, yshift=-4mm] {The metadata\\$M$};
        \node (CDM) [draw, ellipse, right=of Transform, yshift=8mm, xshift=-20mm] {CDM};

        \draw [decorate,decoration={brace,amplitude=12pt}] (-4.2,-0.6) -- (-4.2,-3.1) node[midway]{};

        \node (db) [cylinder, draw, shape border rotate=90, aspect=0.05, right=of I-ETL, align=center] {Database\\with\ CDM\\as schema};

        \draw [->] (I-ETL) -> (db);
        \draw [->] (m1) -> (m2);
    \end{tikzpicture}
}
    \caption{The ETL pipeline for integrating diverse and heterogeneous datasets, keeping interoperability as a first-class citizen.}
    \label{fig:etl}
\end{figure}

Our three-step ETL pipeline leverages the CDM for integrating the input data in the target database, as shown in Figure~\ref{fig:etl}. 
It takes as input the selected datasets $D_1, ..., D_n$ and their accompanying metadata $M$. Next, the three steps Extract, Transform and Load are performed, as detailed in the following. 
While performing them, a number of metrics for assessing interoperability are computed to keep track of interoperability from the start to the very end of the process. 

\vspace{2mm}
\noindent\textbf{Extract step.} The input data, i.e., the metadata $M$ specified by clinical experts and the datasets $D_1, ..., D_n$, is read and stored in memory. No normalization is applied at this stage (postponed to the Transform step). 

\vspace{2mm}
\noindent\textbf{Transform step.} This step creates new objects (instances) that will fit the CDM, i.e., the conceptual model of the target database. While creating the objects, it also ensures that they are interoperable with each other, e.g., by applying interoperability implementation techniques to values. 
We proceed as follows:
\begin{enumerate}
    \item A set of \textsc{Patient} instances is created as the union of all the patients in the input datasets. Each patient is anonymized by assigning an identifier with the scheme $\langle$\texttt{HospitalName:counter}$\rangle$, where \texttt{HospitalName} is the hospital name and \texttt{counter} is an auto-incremented number. No further personal information is stored for patients in the CDM for privacy reasons.
    
    \item Each feature $f \in F$ is created based on the available information in the metadata $M$ and is specialized based on its kind. An identifier of the form $\langle$\texttt{Feature:counter}$\rangle$ is assigned to each of them. 
    If a feature presents both ontology name and code in its metadata, then an \textsc{OntologyResource} is created, respectively with the ontology name and code in attributes \textit{system} and \textit{code}. A query asking for the ontology resource label is sent to the ontology. If a non-empty result is returned, it is stored in the attribute \textit{label}, otherwise the attribute remains empty.
    Next, the attributes \textit{dataType}, \textit{unit} and \textit{visibility} are directly obtained from the metadata. 
    Finally, the attribute \textit{categories} is computed as a map containing pairs of a categorical value (a string) and the associated \textsc{OntologyResource} instance (if it exists).
    
    \item \textsc{Record} instances are created out of the input datasets $D_1, ..., D_n$. For each patient having a non-empty value $v$ for a given feature of kind $k$, a $\bm{k}$\textsc{Record} is created, and the attribute \textit{value} stores $v$. \\    
    The \textit{dataset} attribute is set to the examined dataset $D_i$. A unique identifier is assigned to the new instance, namely an identifier of the form \rev{$\langle$}\texttt{Record:counter}\rev{$\rangle$}. The created record also includes three foreign keys: \textit{register}, \textit{associate}, and \textit{instantiate}, which are identifiers of existing \textsc{Hospital}, \textsc{Patient}, and \textsc{Feature} instances. Specific attributes (e.g., the \textit{baseId} for \textsc{ClinicalRecord} \rev{instances}) are extracted from the dataset while creating the records. \\
    Finally, before setting the \textit{value} with $v$, we make the value as \textbf{interoperable} and \textbf{secure} as possible by proceeding as follows.

    \vspace{2mm}
    \indent\textit{Interoperable.} 
    Each value $v$ undergoes transformations to \textbf{enhance its interoperability}, as detailed in Table~\ref{tab:interoperabilization}. 
    Each technique depends on $v$'s data type; when not applicable, the related functions return the initial value (e.g., string ``tru'' cannot be cast to a boolean, thus the function yields ``tru''). 
    
    \vspace{2mm}
    \indent\textit{Secure.} 
    Then, we \textbf{secure} the interoperable (or initial) value based on the feature \textit{visibility} and \textit{dataType} attributes. When the visibility equals \texttt{anonymize\rev{d}}, \texttt{date} and \texttt{datetime} values are deprived from their day, respectively minutes and seconds. Otherwise (the \textit{visibility} is \texttt{private} or \texttt{public}, or the \textit{dataType} is not \texttt{date} nor \texttt{datetime}), the value $v$ is returned as is. 
    
\end{enumerate}

\begin{table}
    \centering
    \begin{tabular}{ll}
        \hline
        Data type & Function to enable interoperability\\\hline
        \texttt{string} & Normalize $v$ (trim spaces and lower case $v$)  \\
        \texttt{category} & Yield the associated OntologyResource in \textit{categories} \\
        \texttt{date} & Cast $v$ to an ISO format\\ 
        \texttt{datetime} & Cast $v$ to an ISO format\\
        \texttt{boolean} & Cast $v$ to \texttt{True} or \texttt{False} \\
        \texttt{integer}/\texttt{numeric} and $f_\textrm{\textit{unit}} = v_\textrm{\textit{unit}} = \emptyset$ & Cast $v$ to \rev{an} \texttt{integer}, respectively \rev{a} \texttt{numeric}\\
        \texttt{integer}/\texttt{numeric} and $f_\textrm{\textit{unit}} = v_\textrm{\textit{unit}}$ & Remove the unit and cast $v$ (to \texttt{int.}, resp. \texttt{num.})\\
        \texttt{integer}/\texttt{numeric} and $f_\textrm{\textit{unit}} \neq v_\textrm{\textit{unit}}$ & Yield $v$ \\\hline
    \end{tabular}
    \caption{Functions used on Record values to enhance their interoperability. If the technique does not succeed on the input value, the function returns the value as is.}
    \label{tab:interoperabilization}
\end{table}

\vspace{2mm}
\noindent\textbf{Load step.} The last step of the ETL process is to load the instances created in memory into the target database. This includes: the \textsc{Hospital}, \textsc{Patient}, \textsc{Feature}, and \textsc{Record} instances. After insertion, instances are indexed to ensure fast access and querying.

\subsection{Interoperability assessment}
\label{ssec:interop-metrics}

The I-ETL framework finally reports a set of interoperability metrics that are computed across the integration pipeline.
Concerning databases that are created using I-ETL, interoperability can be measured at three levels:
the database is compared with itself (we call this \textit{self-interoperability}); 
with other databases in the same institution (termed  \textit{intra-institution interoperability});
and with the databases of other institutions (termed \textit{inter-institution interoperability}). 
For databases generated through I-ETL that received high metrics scores, we ensure interoperability on all three levels. \rev{These three levels of interoperability can be ensured both from a syntactic and semantic point of view. All data providers use the same data model (the one presented in Section~2.3) and this guarantees syntactic interoperability. For semantic interoperability, we favor it by allowing experts to map the features of interest to well-known and widely adopted ontologies.} 

We derived a set of \textbf{interoperability metrics}, which -respectively- target 
the selected data, 
the created metadata, and 
the ETL process.
Specifically, \textbf{data metrics} are computed to assess the completeness of datasets selected for a specific federated scenario;
\textbf{metadata metrics} assess the coherence and completeness of the metadata created for the selected datasets; and
\textbf{ETL metrics} assess to what extent interoperability and anonymization have been achieved during the ETL process, as well as the coherence of the target database. 

It is crucial that FAIR principles (specifically, interoperability) are put from the beginning of the process 
that achieves the database creation.
Table~\ref{tab:metrics} lists our metrics, which are computed from the start to the end to evaluate self-, intra-institution and inter-institution interoperability. 
The next sections detail the three sets of metrics.

\begin{table}
\begin{tabular}{ll}
\toprule
Step & Metric \\ 
\midrule
\multirow{2}{*}{Data} & (A1) Ratio of selected features \\ 
& (A2) Ratio of datasets that do not require dedicated extraction \\
\midrule
\multirow{5}{*}{Metadata} & (M1) Features with both non-empty ontology \textit{name} and \textit{code} \\ 
& (M2) Features with non-empty \textit{dataType} \\ 
& (M3) Features with non-empty \textit{visibility} \\ 
& (M4) Categorical features with non-empty set of \textit{categories} \\ 
& (M5) Numerical features with non-empty \textit{unit} \\ 
\midrule
\multirow{7}{*}{ETL} & (E1) Presence of non-empty \textit{label} in Ontology Resource\\ 
& (E2) Values for which interoperability implementation has succeeded\\ 
& (E3) Correspondence of numerical values \textit{unit} and Feature \textit{unit} \\ 
& (E4) Presence of categorical value in the Feature \textit{categories} \\ 
& (E5) Records with known Hospital references\\
& (E6) Records with known Patient references\\
& (E7) Records with known Feature references\\\bottomrule
\end{tabular}
\caption{The set of metrics recorded through I-ETL to assess interoperability across the pipeline. \label{tab:metrics}}
\end{table}

\subsubsection{Data metrics}

Data metrics are computed on the selected features $F$ for datasets $D_1, ..., D_n$; we defined two. 
(A1) represents how many features have been retained for the selected datasets, providing insight \rev{of} the extent to which the datasets are mapped to metadata. 
Small subsets of (well-crafted) features are generally preferable in federated scenarios, even if possible information loss must be taken into account. 
(A2) represents the number of datasets that do not require extraction using dedicated tools. 
Dedicated data extraction, defined as the process of identifying and retrieving information using external tools, is often needed with complex data types such as images and genomic files. For instance, MRI scans are images for which dedicated extraction is needed in order to obtain data from the image itself. Examples of data extracted from lung MRI scans include the size of dark regions in the scan, whether abnormalities can be seen, etc. 
In general, using dedicated tools may lead to an overall decrease in interoperability\rev{,} e.g., when different versions or pipelines are used and/or if images are of low quality. 

\subsubsection{Metadata metrics}
For metadata metrics, we evaluate the features $F$ described by the clinicians in the metadata $M$ (Figure~\ref{fig:pipeline}, step 3). 
For all features, we count how many have 
(M1) both \textit{ontology} and \textit{code}, 
(M2) non-empty \textit{dataType}, and (M3) non-empty \textit{visibility}. 
While these three attributes are not mandatory, they are important to achieve high intra- and inter-institution interoperability, crucial for the federated analysis of the datasets within an institution and with other institutions. 

Next, only for categorical features, we count how many of them have at least one associated category (M4). 
Values that cannot be mapped to the specified categories decrease interoperability. 
Finally, we count numerical features that are equipped with a valid \textit{unit} (M5); note that features without this information hinder the interpretation of values (consider, as an example, the ambiguity in determining whether age is expressed in weeks or months). 

\subsubsection{ETL metrics}

ETL metrics are computed during the Transform step while making features and values as interoperable as possible. 
We compute (E1) as the number of ontology resources (whether they represent a Feature or a categorical value) having a non-empty \textit{label}. (E1) captures the interoperability of both features and categorical values because it is of the same importance if one of them carries an empty label (both hinder interoperability). 
The label is empty if the ontology is not provided with a query interface or if the request sent to the ontology fails for any reason. For instance, the \oloinc\ resource, whose code is \texttt{LL4034-6}, does exist at URL \url{https://loinc.org/LL4034-6}, but the query asking for information about this resource returns an empty result, because the \oloinc\ query interface does not access the latest release of the ontology. 
At this point of the ETL, failing queries affect only the OntologyResources' \textit{label}; however, if one wants to retrieve more information for that specific resource, it will not be possible, thus limiting interoperability. 

Next, (E2) counts the ratio of Record values for which interoperability implementation has succeeded (recall Table~\ref{tab:interoperabilization});
(E3) counts the ratio of numerical Record values having the same unit as the Feature they instantiate;
with (E4), we assess the quality of categorical values by checking how many of them were declared in the Feature \textit{categories} list that they instantiate. 
Overall, (E2), (E3), and (E4) measure the coherence between metadata and data.

Finally, (E5), (E6), and (E7) ensure that all created instances refer to (other) objects existing in the database.
This is achieved by counting how many references to hospitals, patients, and features point to instances in the database, preventing broken references.

\subsubsection{Anchoring metrics to FAIR principles}

To legitimate our metrics w.r.t. the FAIR principles, we recall the three sub-principles of Interoperability and explain how our proposed pipeline and metrics conform to them. 

\vspace{2mm}
\noindent
\textit{I1.} \textit{\revII{D}ata \revII{and metadata} use a formal, accessible, shared, and broadly applicable language for knowledge representation.}

\vspace{2mm}
\noindent
Our conceptual common data model is designed in a way such that it can be implemented within any type of database (relational, NoSQL, etc). Moreover, metadata can be easily specified using a tabular file, such as an Excel or CSV file, while following our metadata model. The selected datasets can be of any machine-readable format and dedicated extraction is applied for datasets necessitating pre-processing.  

\vspace{2mm}
\noindent
\textit{I2.} \textit{\revII{D}ata and metadata use vocabularies that follow FAIR principles.}

\vspace{2mm}
\noindent
While specifying metadata, experts are asked to assign an ontology resource to each feature of interest. Despite not mandatory, it is highly recommended in order to achieve good interoperability. Experts are also advised to use well-known and recognized ontologies in their domain. At best, the ontology is publicly accessible; otherwise, it may need to be accessed under authentication, but should not be hospital-dependent. 
The ETL pipeline will reuse these ontologies to encode categorical values, thus increasing interoperability both for metadata and data. This aspect is tracked by ETL metrics assessing ontology resources and categorical values.
    
\vspace{2mm}
\noindent
\textit{I3.} \textit{\revII{D}ata and metadata include qualified references to other data and metadata.}

\vspace{2mm}
\noindent
Each Record instance comprises a reference to a patient, a hospital, and a Feature (qualified references to the database instances), and the dataset from which the value comes (qualified reference to the data). 

\section{Results}
\label{sec:results}

\subsection{Implementation and scenario}


I-ETL is implemented as a Python software built upon v3.12 and the well-known, scalable, and flexible MongoDB~\cite{mongo} database management system (v7.0.12). \rev{The source code is available on GitHub at \url{https://github.com/DEIB-GECO/i-etl}.} It can be easily deployed within medical centers, possibly having various software and hardware setups, employing its Docker image~\cite{docker}.
The software includes a template CSV file for the metadata definition, the ETL, and the computation of interoperability metrics.
It produces the target MongoDB database. \rev{Our software needs to be set up once for each data provider (e.g., each hospital) and can be re-run to add/update data or metadata.}

The selection of relevant datasets and features is left to clinicians. 
To support them in the definition of the features of interest, the metadata model is provided as a pre-defined CSV file whose header corresponds to the 8 attributes of our one-entity metadata model. 
Then, clinicians gather information for each feature of interest and fill the metadata file accordingly. The interoperability score of the target database highly depends on the quality and quantity of the provided metadata.

To showcase and evaluate our approach and its implementation, we consider the study of patients with end-stage kidney disease (ESKD) affected by COVID-19~\cite{kidney-covid-publi}. 
ESKD is the last stage (stage 5) of chronic kidney disease;
this causes kidneys to be at 1/10 of their normal capacity, incapable of effectively removing waste or excess fluid from blood. 
Due to their renal impairment, ESKD patients are at high risk of severe COVID-19, thus necessitating extra care. 
From this study, we consider a scenario where two hospital centers collaborate and share their different kinds of data to run a precise federated analysis. Here, clinicians are interested in the following two questions: 
\begin{itemize}
\item ``Which set of ESKD-related genes favor the development of severe forms of COVID-19?'' 
\item ``Which COVID symptoms are amplified due to the renal impairment of ESKD patients?''.
\end{itemize}

We selected the open and real-life datasets provided by~\cite{kidney-covid-publi} at~\cite{kidney-covid-dataset}, containing 111 patients for which phenotypic, diagnosis, imaging, clinical, and genomic data is provided. We allocate datasets to our two hospitals as follows:
\begin{itemize}
    \item The first hospital $H_1$ processes only genomic data. Genomic data corresponds to RNA sequence counts for a panel of 60,649 genes.
    \item The second one, $H_2$, processes phenotypic, clinical, imaging, and diagnosis data. Clinical data comes from flow cytometry for 45 general cells, while imaging data is about radiology evidence of COVID-19 in patient lungs. Diagnosis data provides information about the possible causes of ESKD as well as its severity.
\end{itemize}

\subsection{Metadata creation}

Starting from the above-mentioned datasets, we defined the set of relevant features $F$ and described them according to our metadata model. 
The metadata has been collected and mapped to existing ontologies manually.

\subsubsection{Hospital 1 metadata}

In Table~\ref{tab:metadata-g}, we report an excerpt of metadata regarding genomic data. Original data contains RNA sequence counts computed for a panel of 60,649 genes for all the patients. In this panel, not all genes are relevant whilst they are costly to process. 
Therefore, we filtered the panel to keep the subsets of genes that are the 1,000 most expressed genes for each patient. 
The reason is that a high sequence count typically indicates that many reads are associated with a gene, suggesting a higher level of gene expression. 
The union of the top 1,000 genes-sets of each patient corresponds to a set of 2,382 genes, each leading to a feature in the metadata. Next, the code associated with each of them corresponds to the gene name without its version (the number after the dot); this has been computed automatically for the 2k genes. All gene codes can be found in the \ohgnc~\cite{hgnc} ontology. In the dataset, values correspond to RNA sequence counts and thus are integers.

\begin{table}
    \centering
    \begin{tabular}{llll}
        \hline
        ontology & code & name & dataType \\\hline
        loinc & 57723-9 & Sample\_ID & string \\
        hgnc & ENSG00000250433 & ENSG00000250433.1 & integer \\ 
        hgnc & ENSG00000258591 & ENSG00000258591.2 & integer \\ 
        hgnc & ENSG00000115902 & ENSG00000115902.11 & integer \\ 
        hgnc & ENSG00000130234 & ENSG00000130234.13 & integer \\ 
        ... & ... & ... & ... \\\hline 
    \end{tabular}
    \caption{The metadata obtained from RNA sequence counts, leading to genomic features. For all features,  \textit{visibility}=\texttt{public}, \textit{unit}=$\emptyset$.}
    \label{tab:metadata-g}
\end{table}

\subsubsection{Hospital 2 metadata}
Metadata for {phenotypic data}, presented in Table~\ref{tab:metadata-p}, has been collected by gathering personal information for each patient. It contains nine variables, each mapped to a \osnomed\ code\footnote{\rev{Tables~\ref{tab:metadata-p}, \ref{tab:metadata-d}, \ref{tab:metadata-i}, and \ref{tab:metadata-c} exhibit \textit{code} values composed of several individual codes, joined with operators such as \texttt{:} or \texttt{=}. The process of joining several codes is referred to as \textit{post-coordination} and is helpful when a concept/variable cannot be described with a single code (mainly because it does not exist in any ontology). We further discuss this point in Section~\ref{ssec:limitations}.}}. 77\% of them are categorical, each of them being associated with up to five categories (column \textit{categories}; each value here has been mapped to a \osnomed\ code). The feature \texttt{calc\_age} is kept private because patient privacy could be violated by looking at young outliers (note that few patients are much younger than the mean age of 68 years old).

\begin{table}
    \centering
    \begin{tabular}{p{2.5cm}llllp{3.3cm}}
        \hline
        code & name & visibility & dataType & unit & categories \\\hline
        422549004 & individual\_id & public & string & - & -\\
        397731000 & ethnicity & public & category & - & asian, white, black, other \\
        734000001 & sex & public & category & - & M, F \\
        397669002 & calc\_age & private & integer & years & -\\
        307294006: & ihd  & public & category & - & no, yes.stent, unknown,  \\
        246454002=``IHD" & & & & & yes.no.intervention, yes.cabg\\ 
        307294006: 246454002=111293003 & previous\_vte & public & category & - & yes.dvt, yes.pe, yes.other, no\\ 
        13645005 & copd & public & category & - & yes, no, copd, bi\\ 
        73211009 & diabetes & public & category & - & yes.T1, yes.T2, no\\ 
        365981007 & smoking & public & category & - & never, not.current, ex \\
         &  &  &  &  & unknown, current\\\hline 
    \end{tabular}
    \caption{The metadata obtained from the patient personal information, leading to phenotypic features. For all features, \textit{ontology}=\osnomed.}
    \label{tab:metadata-p}
\end{table}

For {metadata about diagnoses}, four public features are retained (Table~\ref{tab:metadata-d}) and mapped to a \osnomed\ code. 
The two categorical features have a reasonable number of associated categories, each mapped to a \osnomed\ code. 

\begin{table}
    \centering
    \begin{tabular}{lllp{6.2cm}}
        \hline
        code & name & dataType  & categories \\\hline
        422549004 & individual\_id & string & - \\ 
        46177005 & cause\_eskd & category & DN, Unknown, other, GN, HTN, Genetic\\ 
        405162009: & WHO\_severity & category & moderate, mild, severe, critical \\
        47429007=840539006 &  &  &  \\
        419620001 & fatal\_disease & boolean  & - \\ \hline
    \end{tabular}
    \caption{The diagnosis metadata. For all features,  \textit{ontology}=\osnomed, \textit{visibility}=\texttt{public} and \textit{unit}=$\emptyset$.}
    \label{tab:metadata-d}
\end{table}

{Metadata for imaging data} (Table~\ref{tab:metadata-i}) leads to two features: the patient id and the observed anomaly in the radiology scans (\texttt{radiology\_evidence\_covid}). The latter feature has been associated with ten categories of evidence, each extracted from original imaging scans using dedicated image processing techniques and mapped to a \osnomed\ code. 

\begin{table}
    \centering
    \begin{tabular}{lllp{5.5cm}}
        \hline
        code & name & dataType & categories \\\hline
        422549004 & individual\_id  & string & - \\ 
        840539006: & radiology\_evidence\_covid & category & cxr, no, not.done, yes, yes:CVCX1\\
        363589002= & &  & yes:extensiveBilateralAirspaceConsolidation, \\
        363680008 & &  & yes:extensiveConsolidation, yes:leftBasalInfiltrate, yes:patchyBilateralConsolidation, yes:patchyOpacificationBothLungFields\\\hline 
    \end{tabular}
    \caption{The metadata obtained from the imaging datasets. For all features, \textit{ontology}=\osnomed, \textit{visibility}=\texttt{public} and \textit{unit}=$\emptyset$.}
    \label{tab:metadata-i}
\end{table}

\begin{table}[ht!]
    \centering
    \begin{tabular}{llll}
        \hline
        ontology & code & name & dataType \\\hline
        snomed ct & 422549004 & individual\_id & string \\ 
        loinc & 57723-9 & sample\_id & string \\
        snomed ct & 117400003:260864003=732272000 & CD66b+ $|$ CD45+ & numeric \\ 
        snomed ct & 115412003:260864003=732272000 & CD4+ T $|$ CD45+ & numeric \\ 
         $\emptyset$ & $\emptyset$ & Siglec-1+ $|$ NKG2D-HLA-DR+ & numeric \\ 
        ... & ... & ... & ... \\\hline 
    \end{tabular}
    \caption{The metadata obtained from flow cytometry analyses, leading to clinical features. For all features,  \textit{visibility}=\texttt{public}, \textit{unit}=$\emptyset$.}
    \label{tab:metadata-c}
\end{table}

Finally, {metadata of clinical data} comes from the flow cytometry analyses of patients. It leads to 47 features; a subset of them is shown in Table~\ref{tab:metadata-c}. 
Eight features out of 47 could not be mapped to any \osnomed\ code because one or several of the specified acronyms are not included in the ontology (this can be captured by interoperability metrics, see Section~\ref{ssec:expes-metrics}). This is, for instance, the case of \texttt{Siglec-1} and \texttt{NKG2D}. All the 45 flow cytometry measures lead to numeric values, do not present units, and are all accessible without restriction (\textit{visibility} = \texttt{public}). 

\subsection{ETL execution}
When run on the initial datasets from the two hospitals considered in our scenario, two databases $H_1$ and $H_2$ are consolidated -- see Table~\ref{tab:results-etl} for numbers of corresponding features and records grouped by kind of data.
Both $H_1$ and $H_2$ contain 111 Patients and one Hospital instance. 

\begin{table}
    \centering
    \begin{tabular}{lrrrr}
    \hline
         & $H_1$ & & $H_2$ & \\\hline
         & Feature & Record & Feature & Record \\ \hline
        Phenotypic & - & - & 8 & 888 \\ 
        Clinical & - & - & 45 & 748 \\ 
        Diagnosis & - & - & 3 & 251 \\ 
        Imaging & - & - & 1 & 70 \\ 
        Genomic & 2,382 & 250,103 & - & - \\ \hline
    \end{tabular}
    \caption{Statistics of the databases obtained in experiments.}
    \label{tab:results-etl}
\end{table}

\subsection{Interoperability assessment}
\label{ssec:expes-metrics}

\revII{Finally,} I-ETL reports on the overall interoperability of the target database \revII{by computing and displaying our set of metrics (described in Section~\ref{ssec:interop-metrics}). This final step is highly important to check that obtained databases (as described in Table~\ref{tab:results-etl}) are complete and sound, especially when running federated analyses over several databases.} Table~\ref{tab:expes-interop} lists the scores achieved for \revII{our metrics on} the databases of $H_1$ and $H_2$. 
It also provides the total number of objects accounted for the score. 
The score $s$ ranges from 0 to 1 \revII{(included)}, with higher values indicating better performance. 

In general, we appreciate that I-ETL created two highly interoperable databases from the ESKD-COVID patient data\revII{. In detail, full} interoperability ($s=1.00$) is achieved for \revII{8} metrics in $H_1$ and 10 metrics in $H_2$. 
High interoperability ($0.8\leq s < 1.00$) is achieved for 1 metric in $H_1$ and 2 in $H_2$. 
\rev{L}ow interoperability ($s < 0.8$) is achieved for 2 metrics in $H_1$ and \revII{2} in $H_2$. \revII{Finally, 3 metrics lead to null scores in $H_1$ (N/A values in  Table~\ref{tab:expes-interop}). This is because they were not applicable to the hospital data. For instance, there is no categorical feature for $H_1$, thus (M4) could not been computed. }
By analyzing in more detail the achieved interoperability scores, we observe the following:
\begin{itemize}
    \item \revII{The} ratio of selected features (A1) \revII{is low for} $H_1$ because only 2,382 \revII{genes} were selected among the large panel of 60,649 genes. 
    \revII{This drastic} gene selection was necessary to run I-ETL in a reasonable time, while not relaxing important ones for federated analys\revII{e}s. 
    In $H_2$, a high score is achieved because almost all phenotypic, clinical, diagnosis, and imaging features have been retained. Examples of excluded features are \texttt{WHO\_temp\_severity} (a duplicate of \texttt{WHO\_severity}), \texttt{time\_from\_first\_symptoms} and \texttt{time\_from\_first\_positive\_swab} (the former containing the hour at which COVID symptoms appeared, the latter being the hour at which the nasal test has been done -- both \revII{are} not useful for understanding correlation between ESKD and COVID).
    \item Assessment of (M1) in $H_2$ leads to a score of 0.87 because 8 features \revII{out of 62} could not be mapped to existing ontology concepts. This is, for instance, the case of the last feature shown in Table~\ref{tab:metadata-c} because \texttt{Siglec-1}, \texttt{NKG2D}, \texttt{HLA} and \texttt{DR} are not associated with \revII{any} \osnomed\ code.
    \item (M5) leads to very low scores for \revII{both} $H_1$ and $H_2$, respectively $0$ and $0.02$. This \revII{arises because most of the described features have no associated \textit{unit}} in the metadata, e.g., only \texttt{calc\_age} had a unit in $H_2$. Empty units cover two cases (without distinction): there is no unit for the feature \revII{(as for a ratio)}, and there is one but it has not been specified. To distinguish them and improve interoperability, experts should explicitly specify in the metadata when a feature has no unit. Unfortunately, this \revII{did not} happen \revII{for our experimental datasets}, leading to mostly empty units for numeric features, thus low scores. 
    \item (E1) scores are very high for \revII{both} $H_1$ and $H_2$, meaning that \revII{almost all ontology resources (}associated with feature\revII{s or categorical values)} carr\revII{y} a label. This ensures interoperability and shows that ontologies can provide information about their resources.
    \item (E3) leads to a null score for $H_1$ and a 0-score for $H_2$ for the same reason mentioned above for (M5)\revII{. I}n $H_1$\revII{,} no numeric value has a unit\revII{, thus leading to the N/A score. I}n $H_2$, only \revII{the feature} \texttt{calc\_age} has a unit \revII{specified} in the metadata\revII{,} but no unit was provided in the data \revII{(}as in \revII{the value} ``3 years''\revII{):} thus, the score of 0.
\end{itemize}

\begin{table}
\begin{tabular}{lrrrr}
\hline
Metric & $H_1$ total & $H_1$ score & $H_2$ total & $H_2$ score \\ \hline
A1 & 60,650 & 0.04 \revII{(L)}& 65 & 0.92 \revII{(H)}\\ 
A2 & 5 & 1.00 \revII{(F)}& 5 & 1.00 \revII{(F)}\\
M1 & 2,382 & 1.00 \revII{(F)}& 62 & 0.87 \revII{(H)}\\ 
M2 & 2,382 & 1.00 \revII{(F)}& 62 & 1.00 \revII{(F)}\\ 
M3 & 2,382 & 1.00 \revII{(F)}& 62 & 1.00 \revII{(F)}\\ 
M4 & \revII{N/A} & \revII{N/A} & 10 & 1.00 \revII{(F)}\\ 
M5 & 2,382 & 0.00 \revII{(L)}& 46 & 0.02 \revII{(L)}\\ 
E1 & 2,382 & 0.99 \revII{(H)}& 46 & 1.00 \revII{(F)}\\ 
E2 & 250,103 & 1.00 \revII{(F)}& 1,957 & 1.00 \revII{(F)}\\ 
E3 & \revII{N/A} & \revII{N/A} & 1 & 0.00 \revII{(L)}\\ 
E4 & \revII{N/A} & \revII{N/A} & 1,028 & 1.00 \revII{(F)}\\ 
E5 & 250,103 & 1.00 \revII{(F)}& 1,957 & 1.00 \revII{(F)}\\
E6 & 250,103 & 1.00 \revII{(F)}& 1,957 & 1.00 \revII{(F)}\\
E7 & 250,103 & 1.00 \revII{(F)}& 1,957 & 1.00 \revII{(F)}\\\hline
\end{tabular}
\caption{\revII{I}nteroperability \revII{a}ssessment for the databases located in $H_1$ and $H_2$. \revII{Interoperability levels are: full (F: $s=1$), high (H: $0.8\leq s < 1$), low (L: $s<0.8$). } \label{tab:expes-interop}}
\end{table}


\section{Discussion}
\label{sec:discussion}

\subsection{Challenges and limitations}
\label{ssec:limitations}

The primary challenge we faced was to design a conceptual data model that could fit the various kinds of data brought by hospitals and clinical centers. Reusing existing CDMs was deemed not possible, \rev{because many of them} are tied to \rev{entities of particular use cases}, e.g., OMOP~\cite{omop} allows to represent observational data and is hardly exten\rev{si}ble to model genomic information. \rev{The more general ones, e.g., FHIR~\cite{fhir}, \revII{lift the above limitation by exhibiting entities of various kinds}. \revII{This makes them well suited to initiatives where new data needs to be collected, processed, and stored.} However, they are not yet general enough to design automatic integration workflows \revII{for existing data} (\revII{as opposed to} hand-made ETL pipelines)\revII{, notably due to they reliance on specific attributes}}. \rev{Following those observations}, we \rev{propose} a \rev{novel common data} model based on the notions of features and records -- abstract concepts of how any dataset can be represented. It currently represent\rev{s} six kinds of healthcare-related data, but \rev{is} easily usable with other kinds, e.g., administrative or surgery-related data. \rev{This makes our conceptual model general enough to be used as a CDM in a wide variety of healthcare projects. In a broader scope, our framework could be utilized in many other contexts, e.g., journalistic sources, spatial databases or social human sciences sources, while only requiring to design a new CDM (such as the one presented in Figure~\ref{fig:conceptual-model}) reflecting entities of the domain and leveraging the notions of feature and record}.



For what concerns metadata creation -- a crucial step to achieve high interoperability -- the main challenge lies in the contribution of clinicians, who often do not have the time and/or knowledge to create it. 
So far, experts need to manually \revII{define all the features they are interested in, specify their related information and} map each \revII{of them} to an ontology code. \revII{Creating metadata may represent considerable manual work, especially for federated analyses where several datasets are joined. Nonetheless, this is the only part where experts are required to do a technical work, supported by our easy-to-fill metadata model.} Even though each ontology is tailored to a particular type of healthcare data, e.g., \ohgnc\ is for genes and \oloinc\ is for clinical measurements, 
finding appropriate ontologies and then searching them for suitable concepts is very time-consuming. 
\revII{Also}, some concepts are \rev{very specific, thus are} not represented in any well-known -commonly adopted- ontology\rev{. They can} be created \rev{through \textit{post-coordination}, a process to} join several exiting codes. \rev{For instance, the feature \texttt{previous\_vte} (whether the patient already had a venous thromboembolism) does not exist in \osnomed\ but can be represented with the following association of codes: ``307294006:
246454002=111293003'' (meaning that there is an occurrence of venous thrombosis in the patient's personal history). Creating post-coordinated codes is} even more time-consuming. 

Manual mapping also suffers from being error-prone, especially when the number of features is large. 
To limit experts' manual efforts and errors, we envision semi-automatic support that 
($i$) proposes a set of ontologies in which the concept is likely to appear (e.g., with BioPortal Recommender~\cite{bioportal-recommender,martinez2017ncbo}); 
($ii$) lets experts select the most appropriate one; 
($iii$) automatically proposes a set of codes that fit the concept in the selected ontology (e.g., by integrating BioPortal Search~\cite{bioportal-search,whetzel2011bioportal}); and 
($iv$) lets the expert select the most appropriate term code. 
Such methods should be used in a human-in-the-loop process.
Indeed, they do rely on various metrics, including semantic similarity measures, but have very vague or no context about the scenario, thus may return inappropriate codes. 
Moreover, it is crucial that medical experts are provided user-friendly support to share their knowledge on the context, so that accurate domain-specific information can be ensured. 

Concerning FAIR principles, the sub-principle I3, stating that qualified references to the data and metadata are necessary, is only partially implemented so far. Indeed, qualified references to the data are already included because each entity in the conceptual data model has an \textit{identifier}. However, qualified references to the metadata are not yet included, but will be in subsequent work by providing a catalog to browse and search datasets based on their metadata.

\subsection{Outlook}

Our I-ETL framework has been developed in the context of a large European project called BETTER~\cite{better}, whose overarching objective is to develop a decentralized and federated analysis of healthcare data. 
In this project, seven clinical centers are involved and \revII{they all work on the general domain of genetic rare diseases. Yet, they derived} three use cases \revII{of interest}, namely, pediatric intellectual disability, retinal dystrophies, and self-harm behaviors for autistic patients. \revII{While all of them rely on genomic data (at least), they also use different kinds and forms of data, thus highlighting the need for a general and easy-to-use framework to integrate and process them. In practice, e}ach center provides datasets from a plethora of different kinds \rev{for the use case they are interested in. Starting from this, we discussed with them the healthcare research questions raised by their use cases, their available data (clinical measurements, genomic variants, MRI scans, etc.) and their ideas in terms of Federated Learning analyses. Next, we designed I-ETL and our two conceptual models (for metadata and data). In parallel, clinical experts discussed the metadata to be considered and filled out the metadata for each of their datasets by leveraging our metadata model. At this stage, hospitals have agreed on common and specific features to include in the metadata. By doing so, they ensure that their databases can be joined for further analyses (otherwise, each hospital would end up with a unique feature set).} \revII{This is where most of the work happens for medical experts (formulate questions, find datasets, specify metadata); the rest of the pipeline is automatic and leads to a ready-to-use database.}

\rev{We are currently deploying our framework inside each partner hospital and collecting feedback on this deployment as well as the usage of our tool. The I-ETL pipeline has been well-received by all the different stockholders involved in the BETTER project. Even if the overhead in the data integration pipeline is costly and demanding, all the actors found that the overall process of creating an interoperable database on their server is worth the effort as long as it allows them to later create AI federated algorithms for medical decision making. Our next task is to discuss with them to finalize their FL scenarios and implement corresponding algorithms.}

As future developments, BETTER aims at providing 
($i$)~a catalog for browsing metadata and aggregated data of target databases, as well as ($ii$)~a platform for running decentralized and federated analyses of the data. 

The catalog will be a website listing all the accessible databases and providing aggregated views of the data for each of them \rev{-- we already initiated this work in~\cite{profiling-better}}. For instance, the clinicians of the BETTER project may browse the metadata of different hospitals to check which other institutions they can join forces with. 
They may also take a deeper look at the aggregated data (while original data and the target database are never accessible outside of centers). For instance, they can investigate the patient age distribution as well as the set of diseases of patients of another institution to understand whether a federated analysis combining their data would make sense. 

After deciding which datasets and which institutions can be joined, federated analysis will be run on a platform based on the Personal Health Train (PHT~\cite{pht}) paradigm. 
This platform will include statistical and AI-based models for analyzing various data stored in the underlying I-ETL-based databases. 
In the end, clinicians will be able to explore the results of the federated computations and gain insights toward solving their research healthcare questions.

\section{Conclusions}
\label{sec:conclusions}

In this paper, we presented I-ETL, a framework for integrating heterogeneous healthcare datasets with interoperability as a first-class citizen. Our contributions are the following. First, we proposed a general data model for a large set of health datasets, including clinical, phenotypic, genomic, diagnosis, imaging, and medication data. This conceptual model \rev{serves as a common data model for various healthcare settings. Its main strengths are to take into consideration experts' knowledge (metadata) and to be} easily exten\rev{si}ble\rev{/}tunable \rev{for other} scenarios. Next, we proposed and implemented an ETL pipeline for transforming the input data into a database designed on our conceptual model. \rev{Incidentally, I-ETL also allows for resource savings (personnel and servers) because it is easy to put in place and does not require a large-scale centralized server.} 
Finally, I-ETL provides a set of across-pipeline metrics for assessing the interoperability level throughout the whole process of integrating the input data into a target database. \rev{Ensuring and assessing interoperability also goes into the direction of data quality; well-conceptualized and homogenized datasets will be easily used for FL analyses.} Experiments on a small open-data-based scenario with two hospitals have shown that I-ETL can achieve high interoperability scores, thereby enabling effective collaboration between different medical centers, notably via federated analysis of the target databases. 

Several research directions \rev{arise from the present work -- some of which are already ongoing. First, we are now working on the querying of the interoperable databases (available at each center) through the catalog. This task is complex because, for privacy reasons, the catalog relies on aggregated data only and the real data in the hospital servers cannot be accessed. Therefore, the challenge here is to find the right balance between super-aggregated data (very safe but not very useful due to the high information loss) and low-aggregated data (more useful but with privacy concerns). A subsequent direction is the design of a human-in-the-loop recommendation module for metadata. This would} automatic\rev{ally} recommend ontologies \rev{and} codes \rev{for a given set of features (recall Section~\ref{ssec:limitations}), allowing experts to save time and reduce errors while keeping control of the obtained metadata. \rev{Another interesting addition would be to add more context to the records, e.g., to know whether a value has been observed before or after surgery. This would contribute to a richer common data model while remaining as general as possible.} In parallel with these three directions, the BETTER partners work on} the implementation of the federated analysis platform to enable the design and secure execution of Federated Learning tasks. 

\rev{With this project, we learned that there is no ``one-size-fits-all'' solution, especially when working in large consortium and projects. Despite these challenges, bringing computer science methods and developments to the healthcare sector opens the road to better health systems, improving citizens' global health.}

\backmatter

\bmhead{List of abbreviations}
\hfill \break \noindent
CDM: Common Data Model\\
CSV: Comma-Separated Value \\
DICOM: Digital Imaging and Communications in Medicine\\
ETL: Extract-Transform-Load\\
FAIR: Findable, Accessible, Interoperable, Reusable\\
\ohgnc: HUGO Gene Nomenclature Committee \\
\oloinc: Logical Observation Identifiers Names and Codes\\
\oomim: Online Mendelian Inheritance in Man\\
PHT: Personal Health Train\\
\osnomed: SNOMED Clinical Terms\\
VCF: Variant Call Format

\section*{Declarations}

\bmhead{Ethics approval and consent to participate}
Not applicable.

\bmhead{Consent for publication}
Not applicable.

\bmhead{Data availability}
The datasets used in this manuscript are available in the Zenodo repository at \url{https://zenodo.org/records/7410194} in open source.

\bmhead{Competing interests}
The authors declare that there are no conflicts of interest.

\bmhead{Funding}
This work is supported by the Horizon Europe project BETTER, Grant agreement n. 101136262. 

\bmhead{Authors' contributions}
NB, AB, BB, and PP conceived the research;
NB and BB jointly conceptualized the framework;
NB designed/implemented the ETL pipeline and software and performed the experiments;
BB curated the ontology mappings;
AB and PP acquired funding;
PP supervised the project;
NB drafted the manuscript;
AB, BB, and PP revised/edited the manuscript.
    

\bmhead{Acknowledgments}
We thank all the partners involved in the BETTER project for their valuable contributions and feedback.


\bibliography{sn-bibliography}

\end{document}